\documentstyle[aaspp4,amstex,psfig,epsfig,12pt]{article}
\def\gms{{\rm\,g\,s^{-1}}}
\def\pc{{\rm\,pc}}
\def\simlt{\lower.5ex\hbox{$\; \buildrel < \over \sim \;$}}

\lefthead{Fromerth and Melia}
\righthead{Shock Formation of BLR clouds}

\begin{document}

\title{The Formation of Broad Line Clouds in the Accretion Shocks of Active Galactic Nuclei}
\author{Michael J. Fromerth\altaffilmark{1}}
\affil{Department of Physics, The University of Arizona, Tucson, AZ 85721}
\and
\author{Fulvio Melia\altaffilmark{2}}
\affil{Department of Physics and Steward Observatory, The University of Arizona, Tucson, AZ 85721}
\altaffiltext{1}{NSF Graduate Fellow.}
\altaffiltext{2}{Sir Thomas Lyle Fellow and Miegunyah Fellow.}
\begin{abstract}

Recent work on the gas dynamics in the Galactic Center has
improved our understanding of the accretion processes in
galactic nuclei, particularly with regard to properties such
as the specific angular momentum distribution, density, and
temperature of the inflowing plasma.  With the appropriate
extrapolation of the physical conditions, this information can
be valuable in trying to determine the origin of the Broad
Line Region (BLR) in Active Galactic Nuclei (AGNs). In
this paper, we explore various scenarios for the cloud
formation based on the underlying principle that the source
of plasma is ultimately that portion of the gas trapped
by the central black hole from the interstellar medium.
Based on what we know about the Galactic Center, it is
likely that in highly dynamic environments such as this,
the supply of matter is due mostly to stellar winds from
the central cluster.

Winds accreting onto a central black hole are subjected to
several disturbances capable of producing shocks, including
a Bondi-Hoyle flow, stellar wind-wind collisions, and turbulence.
Shocked gas is initially compressed and heated out of thermal
equilibrium with the ambient radiation field; a cooling
instability sets in as the gas is cooled via inverse-Compton
and bremsstrahlung processes.  If the cooling time is less than
the dynamical flow time through the shock region, the gas may
clump to form the clouds responsible for broad line emission
seen in many AGN spectra.  Clouds produced by this process display
the correct range of densities and velocity fields seen
in broad emission lines.  Very importantly, the cloud distribution agrees
with the results of reverberation studies, in which it is seen that
the central line peak (due to infalling gas at large radii) responds
slower to continuum changes than the line wings, which originate in
the faster moving, circularized clouds at smaller radii.

\end{abstract}


\section{INTRODUCTION}

The spectra of many AGNs, including Seyfert galaxies
and quasars, are distinguished by strong, broad emission lines,
with a full width at half maximum intensity (FWHM) of
$\sim 5000\ \mathrm{km\ s^{-1}}$, and a full width at zero intensity (FWZI) of
 $\sim 20,000\ \mathrm{km\ s^{-1}}$ (e.g., \cite{BMP97}).
From the observed strength of UV emission lines, we know that the temperature
of the emitting plasma is on the order of a few times $10^4\ \mathrm{K}$ 
(e.g., \cite{DEO89}), insufficient to produce the observed line widths
via thermal (Doppler) broadening.  Instead, bulk motions of the BLR gases
appear to be responsible for the line broadening.

Because reverberation studies show a direct response of emission-line strengths 
to continuum variability (e.g., \cite{JC91}), we know that the BLR gas must be
photoionized by the continuum.  The International AGN Watch consortium has 
carried out long-term optical and ultraviolet monitoring on a set of four 
Seyfert~1 galaxies:  NGC~5548 (e.g., \cite{KTK95}; \cite{BMP99}), NGC~3783 
(\cite{GAR94}; \cite{GMS94}), Fairall~9 (\cite{PMRP97}; \cite{MSL97}), 
and NGC~7469 (\cite{IW97}; \cite{SJC98}); and the broad line radio galaxy 
3C~390.3 (\cite{MD98}; \cite{PTOB98}).  In all sources, it is observed that 
higher ionization lines respond faster than lower ionization lines.
This would indicate that the former are found at smaller radii using a 
simple $r \sim c\, \tau_{delay} $ argument.  The response time for the same 
line varies by source, even when the luminosities are very similar, indicating 
that the simple $r_{\mathrm{BLR}} \propto L^{1/2}$ rule alone does not 
determine the size of the BLR.  Indeed, other factors, such as geometry, 
viewing angle, and spectral energy distribution (SED) may play equally 
important roles in determining its volume (e.g., \cite{AR95}; \cite{AW97}).

What the reverberation studies do tell us, however, is that the size of the 
BLR ranges from a few to several hundred light days, with a radial dependence 
on ionization state and possibly other physical properties.
It is also seen that the response delay within a given source tends to increase 
with increasing luminosity (\cite{BMP99}), consistent with the $r_{\mathrm{BLR}} 
\propto \mathit{L^{1/2}}$ rule.  The optical continuum displays little or no lag 
($\tau_{delay} \lesssim 2$ days) in variability with respect to the ultraviolet 
continuum, and the amplitude of variations are typically weaker at longer 
wavelengths (\cite{SJC98}).

The ionization parameter, $U \equiv n_{\gamma} / n$, which is the ratio of the 
number density of hydrogen ionizing photons $n_{\gamma}$ to the number density of 
hydrogen nuclei $n$, determines the physical state of the BLR plasma.
In modeling the BLR with photoionization simulations, it is observed that a value 
of $10^{-2} < U < 1$ (e.g., \cite{HN90}) is required to reproduce the correct line 
strength and ionization state of the gas.
Because of the radial structure of the BLR, it is likely that the density of 
emitting gas varies with radius, within the range $10^8\ {\mathrm cm}^{-3} 
\lesssim n \lesssim 10^{11}\ {\mathrm{cm}^{-3}}$ (e.g., \cite{BMP97}). However,
Marziani et al. (1996) find that the BLR may extend to even higher densities
($n_e \sim 10^{12.5} \;\mathrm{cm}^{-3}$).
In addition, the BLR may contain a mixture of optically thick and thin gases.
Optically thick gas must be present to account for the variability of the low 
ionization \ion{Mg}{2}, Ly$\alpha$, and Balmer lines (e.g., \cite{GJF92}).
Optically thin gas, on the other hand, may account for the Baldwin effect, a 
negative correlation between the ultraviolet emission-line equivalent width 
and continuum luminosity; and the Wamsteker-Colina effect, a negative correlation 
between \ion{C}{4} $\lambda 1549$/Ly$\alpha$ ratio and continuum luminosity 
(\cite{JCS95}). Green (1995), however, suggests that these effects are due to 
changes in the SED with luminosity.  The set of other important observational 
constraints include (i) the absence of a deep Ly$\alpha$ absorption edge in 
AGN spectra indicates that the BLR gas must cover only a small fraction 
(5 - 25\%) of the continuum source (e.g., \cite{MB97}), and (ii)
the observed line strength to continuum ratio requires a small volume 
filling factor ($\sim 10^{-7}$; \cite{HN90}).

The recent work in modeling (and observations of) the BLR suggests that the 
clouds are spread over a wide range of radii and may display a wide range of particle 
density, $n_c$, at each radius.  The emission from each cloud can be determined 
largely from the ionization parameter $U$.  An important result of this 
``locally optimally emitting cloud" model is that the predicted integrated 
spectrum from all clouds depends only weakly on the parameters, including the 
SED, column density of the clouds, and the cloud distribution as a function of 
radius (\cite{JB97}).  However, the spectrum does depend strongly on $n_c$ and 
on the elemental abundances.  A consequence of the weak dependence on many of 
the input parameters is that several different models may be able to account for 
at least some of the observed spectra.

\subsection{A Sample of Current Models}

There are several models in the literature that account for the origin and nature of the 
BLR.  For example, Emmering, Blandford, \& Shlosman (1992) propose that the BLR is 
associated with magnetohydrodynamic winds originating in a dusty molecular accretion disk.
Dense molecular clouds are loaded onto magnetic lines threading the disk and are 
centrifugally accelerated outward.  Being exposed to the central continuum, these clouds 
are quickly photoionized and produce the observed emission lines.  This model correctly 
``postdicts'' both the observed shape and differential response time of the 
\ion{C}{4} $\lambda 1549$ line, with its mid-red wing portion responding 
fastest to continuum variations (\cite{MB97}).

In a different model, Murray et al. (1995) propose that the broad absorption lines 
(BALs) seen in $\sim 10 \%$ of radio quiet QSOs are produced in outwardly flowing 
radiation- and gas-pressure driven winds rising from an accretion disk.
These winds would also be partially responsible for the broad line emission.
As this model requires shielding of the absorbing gas from soft X-rays, the
absence of BALs in radio-loud quasars and Seyfert galaxies is explained by the 
fact that these objects are strong X-ray emitters (\cite{MC95}).  Cassidy 
\& Raine (1996) present a similar model in which BLR clouds form as the result 
of the interaction of an outflowing wind with the surface of an accretion disk.

Alexander \& Netzer (1994) propose that the AGN broad-line emission originates 
in the winds or envelopes of bloated stars in the nuclear environment.
They obtain good agreement with the line ratios and response features seen in AGNs, 
but they encounter some difficulty in reproducing the broad line wings (\cite{AN97}).

Finally, Perry \& Dyson (1985) propose that the BLR clouds are formed as the result 
of a cooling instability that occurs when an outflowing wind from the black hole 
encounters an ``astrophysical obstacle'' and is shocked.  Rapid cooling in the 
shocked gas causes the plasma to clump into clouds, and the observed line widths 
are then due to cloud acceleration along the shocks.

\bigskip
\subsection{The Analogy with the Galacic Center}
\label{sub:AGC}
\medskip
In this paper, we will take the approach that it may be worthwhile
in formulating a model for the BLR to seek guidance from the galactic
nucleus we know best---that of our own Galaxy.  The evidence for the
presence of a supermassive black hole, coincident with the radio source
Sgr A* at the Galactic Center, is now the most compelling of any such
systems (for a recent review dealing mostly with the observational
characteristics of this region, see Mezger, Duschl \& Zylka 1996;
for a summary of the theoretical status concerning Sgr A*, see
Melia 1998). The motions of stars within 1 pc of Sgr A* seem to require
a central dark mass of $(2.61\pm 0.35)\times 10^6\;M_\odot$
(Genzel et al. 1997; Ghez et al. 1998), in good agreement with earlier
ionized gas kinematics and velocity dispersion measurements.

Our proximity to the Galactic Center provides us with the rather unique
opportunity of examining the gas dynamics surrounding such a massive
point-like object with unprecedented detail.  Combined with multi-dimensional
hydrodynamical simulations, this extensive body of multi-wavelength data
is opening our view into the complex patterns of plasma-plasma and
plasma-stellar interactions.  It is likely that many of Sgr A*'s
characteristics are associated with the liberation of gravitational energy
as gas from the ambient medium falls into a central potential well
(Melia 1994; Ruffert \& Melia 1994).  There is ample observational
evidence in this region
for the existence of rather strong winds in and around Sgr A* itself
(from which the latter is accreting), e.g., the cluster of mass-losing,
blue, luminous stars comprising the IRS 16 assemblage located within
several arcseconds from the nucleus.  Measurements of high outflow
velocities associated with IR sources in Sgr A West (Krabbe et al. 1991)
and in IRS 16 (Geballe et al. 1991), the $H_2$ emission
in the circumnuclear disk (CND) from molecular gas being shocked by a
nuclear mass outflow (Genzel et al. 1996; but see Jackson et al. 1993 for the
potential importance of UV photodissociation in promoting this $H_2$
emission), broad Br$\alpha$, Br$\gamma$ and He I emission lines from
the vicinity of IRS 16 (Hall et al. 1982; Allen et al. 1990), and
radio continuum observations of IRS 7 (Yusef-Zadeh \& Melia 1992),
provide clear evidence of a hypersonic wind, with a velocity
$v_w \sim500-1000$ km s$^{-1}$, a number density $n_w\sim10^{3-4}$
cm$^{-3}$, and a total mass loss rate $\dot M_w\sim3-4\times10^{-3}\;
M_\odot\ {\mathrm yr^{-1}}$, pervading the inner parsec of the Galaxy.

In recent years, several studies have addressed the question of what
the physical state of this gas is likely to be as it descends into the
deepening gravitational potential well of the massive black hole.
In the classical Bondi-Hoyle (BH) scenario (Bondi \& Hoyle
1944), the mass accretion rate for a uniform hypersonic flow is
$\dot M = \pi {R_A}^2 m_H n_w v_w$, in terms of the accretion
radius $R_A \equiv 2 G M / {v_w}^2$.  With the conditions at the
Galactic Center (see above), we would therefore expect an accretion rate
$\dot M \sim 10^{22} \gms$ onto the black hole, with a capture
radius $R_A \sim 0.02 \pc$.

In reality the flow past the supermassive black hole is not likely to be
uniform, so this value of $\dot M$ may be greatly underestimated.
For example, one might expect many shocks to form as a result of wind-wind
collisions within the cluster of wind producing stars, even before the
plasma reaches $R_A$.  With this consequent loss of bulk kinetic energy,
it would not be surprising to see the black hole accrete at an even larger
rate than in the uniform case.  The implications for the gas dynamics
in the region surrounding the black hole are significant.  Coker \& Melia
(1997) have undertaken the task of simulating the BH accretion from the
spherical winds of a distribution of 10 individual point sources located
at an average distance of a few $R_A$ from the central object. The results
of these simulations show that the accretion rate depends not only on the
distance of the mass-losing star cluster from the accretor but also on the
relative spatial distribution of the sources.

These calculations indicate that to fully appreciate the morphology
of the gaseous environment surrounding the accretor, one must pay
particular attention to the spatial distribution of specific angular
momentum $l$ in the accreting gas.  Written as $l\equiv \lambda
c R_S$, where $R_S\equiv 2GM/c^2$ is the Schwarzschild radius,
the accreted $\lambda$ can vary by $50\%$ over $\simlt 200$ years
with an average equilibrium value of $10-50$ for the conditions in
the Galactic Center. This is interesting in view of the fact that
earlier simulations based on a uniform flow---the ``classic'' Bondi-Hoyle
accretion, producing a bow shock---resulted in $\langle\lambda\rangle
\sim 3-20$. It appears that even with a large amount of angular
momentum present in the wind, relatively little specific angular momentum
is actually accreted.  This is understandable since clumps of gas with a
high specific angular momentum do not penetrate to within $1\ R_A$.
The variability in the sign of the components of $\lambda$ suggests
that if an accretion disk forms at all, it dissolves and reforms (perhaps)
with a different sense of spin on a time scale of $\sim 100$ years (\cite{RFC97}).

The fact that AGNs are significantly more gas rich and display a more
powerful array of phenomena than the Galactic Center could mean that these
ideas derived from the latter may not be valid in the case of
the former. But one area where this type of gas morphology would certainly
have a significant impact is in the structure and nature of the BLR.
Our intention in this paper is therefore to frame our investigation
of the BLR in AGNs with the conditions (i.e., clumping, distribution in
specific angular momentum $\lambda$, density and temperature) we now
believe to be prevalent in the Galactic nucleus, though scaled accordingly.

\section{OVERVIEW OF THE MODEL}

\subsection{An Accretion Shock Scenario for the Production of BLR Clouds}

We suggest that many of the observed properties of the BLR can be explained
by a simple picture of cloud production within the accretion shocks
surrounding the central black hole.  For this, we shall adopt several of the 
ideas introduced in Perry \& Dyson (1985, hereafter PD85) for the formation 
of clouds from cooling instabilities in these regions.  In their model, a 
hypersonic, outflowing wind is incident upon a supernova remnant or other 
astrophysical obstacles, causing bow shocks to form around them.  The shocked 
gas is compressed and heated, and is brought out of thermal equilibrium with 
the radiation field.

The equilibrium temperature $T_{eq}$ of a gas whose heating/cooling is dominated 
by radiative processes is determined by another ionization parameter (\cite{KMT81})
\begin{equation}
\Xi \equiv \frac{F_{ion}}{n k T c}\;,
\label{eq:Xi}
\end{equation}
where $F_{ion}$ is the ionizing flux between 1 and $10^3$ ryd; then,
$T_{eq} = T_{eq}(\Xi)$.  
Because the shock temperature $T_s \gg T_{eq}$, the shocked gas will rapidly
cool via inverse-Compton and bremsstrahlung processes.  If the cooling time
is shorter than the dynamical time for the gas to flow along the shock, the
cooled gas will clump and form clouds, which then stream along and behind
the shock.

An important parameter in determining $T_{eq}$ is the Compton temperature $T_C$, 
at which Compton heating and cooling processes balance (e.g., \cite{KMT81}; \cite{PWG86}).
The Compton temperature is highly sensitive to the shape of the continuum, 
particularly at high (X-ray and gamma ray) energies; typical AGN spectra 
have $T_7 \approx 0.01 - 5$ (\cite{MF87}), where $T_7 \equiv T_C/ (10^7\ {\mathrm{K}})$ .
For these high values of $T_C$, we can state a couple of generalities about $T_{eq}$:
if $\Xi \gg 1$, Compton processes dominate, and $T_{eq} \simeq T_C$;
if $\Xi \ll 1$, collisional (bremsstrahlung) and recombination-line cooling 
dominate, and $T_{eq} \simeq 1-3 \times 10^4$ K.

In our picture, we assume that gravitation dominates over the outward radiation
pressure within the BLR (either because the outward radiation field is sub-Eddington,
or because the radiative emission is anisotropic), allowing a hypersonic, accreting 
wind to feed the central black hole.  In the AGN context, we re-examine the PD85 
result for stellar wind bow shocks, and extend the idea of ``astrophysical obstacles''
to include Bondi-Hoyle accretion shocks and density perturbations due to wind-wind 
collisions and turbulence in the accretion flow.  As discussed above, this is motivated
by the recent simulations of the highly variable gas flows at the Galactic Center.

\subsection{Model Parameters}
\label{sub:MP}

Our model requires the specification of several parameters, including the density,
velocity, and temperature profiles of the accreting wind as functions
of radius, as well as the intensity and Compton temperature of the continuum.
In order to keep our arguments general, we choose to model the wind
flow using simple dimensional requirements.  We adopt the view that the
accreting gas circularizes before it reaches the event horizon, thereby
forming a disk at small radii. The existence of an accretion disk in AGNs
is inferred from, e.g., the axisymmetry observed in many sources 
(\cite{MSB96}; \cite{JG94}).  Again writing the specific angular
momentum as $l \equiv \lambda c R_S$, it is easy to show that the 
circularization radius is 
\begin{equation}
r_{circ} = 2 \lambda^2 R_S \ .
\end{equation}

If the central object dominates the gravitational
potential, then the characteristic wind
velocity scales as the free-fall velocity, $v_w(r) \approx v_{ff} = \sqrt{2GM/r}$.  
Due to mass conservation (which in the outer region gives $\dot{M} = 
4 \pi r^2 n_w v_w$) and assuming that the bolometric luminosity is related 
to the mass accretion rate by
\begin{equation}
L_{bol} = \epsilon \dot{M} c^2 \ ,
\label{eq:L_bol}
\end{equation}
where $\epsilon < 1$ is the accretion efficiency, the wind density can be expressed as
\begin{equation}
n_w(r) = \frac{L_{bol}}{4 \pi\, \sqrt{2GM}\, \epsilon\, m_H\, c^2\, r^{3/2}} \ .
\label{eq:n_w}
\end{equation}

The temperature $T_w$ of the gas undergoing steady,
spherical infall is described by the Equation (\cite{MF87})
\begin{equation}
        \frac{dT_w}{dr} = -\frac{T_w}{r} + \frac{F_c(T_w)}{v_w} \;, 
\label{eq:T(r)}
\end{equation}
which we use as an approximation for our non-steady flow.
In Equation~(\ref{eq:T(r)}), the first term represents compressional
heating and the second term is due to radiative heating/cooling, for
which $F_c(T)$ is the cooling function of the gas (c.f., Eq.~(\ref{eq:F_c})
in the Appendix).  As we are mainly concerned with establishing a 
\emph{minimum} temperature of the flow, we have neglected the effects of
viscosity, shocks, and the dissipation of magnetic energy, all of which may 
raise the value of $T_w$.

In a recent study, Wandel, Peterson and Malkan (1999) used reverberation 
data to infer central masses of $M_8 \approx 0.02 - 4$ (where $M_8 \equiv 
M/(10^8\ M_{\odot})$) for a sample of 17 Seyfert~1 galaxies and 2 quasars.
They also determined a mass -- monochromatic luminosity relation of 
$L(5100\ {\rm \AA}) \simeq 10^{44}\, M_8^{1.25}\ \mathrm{erg\ s^{-1}}$ for 
these objects.  Setting $L_{bol} = f_{bol} L(5100\ {\rm \AA})$ for the
monochromatic -- bolometric luminosity relation, with $f_{bol} = 10$ 
consistent with the findings of Bechtold et al. (1987), we obtain
\begin{equation}
L_{bol} = 10^{45} M_8^{1.25}\ \mathrm{erg\ s^{-1}} \ .
\label{eq:L_bol2}
\end{equation}
For the ionizing luminosity, we set $L_{ion} = f_{ion} L(5100\ {\rm \AA})$, 
where $f_{ion} = 2.5$ has been chosen as a fiducial value.
Under these assumptions, the only free parameters in the model are 
$M$, $\epsilon$, $T_C$, and $\lambda$.

\subsection{Optical Depth of the Flow}
The observed absence of the Fe K-shell edge in most AGN spectra indicates 
that the inter-cloud medium must be optically thin to X-radiation 
(e.g., \cite{MF87}).  This limit is written as $\tau_K < 1$, where
\begin{equation}
\tau_K = \int_{r_{circ}}^{\infty} \delta_{Fe}\ \sigma_K n_w(r)\, dr
\label{eq:tauK}
\end{equation}
is the Fe K-shell optical depth, $\sigma_K = 2.3 \times 10^{-20}\ 
\mathrm{cm^2}$ is the total K-shell cross section (e.g., \cite{MM83}), 
and $\delta_{Fe}$ is the elemental abundance of iron.
Assuming that $\delta_{Fe} = 3.3 \times 10^{-5}$ (corresponding to the 
local ISM value; \cite{Dalgarno87}) and that all Fe ions in the flow 
retain at least two electrons (a conservative estimate given the 
likely high temperature of the gas), then the condition for the flow 
to remain optically thin to X-radiation is
\begin{equation}
\lambda \gtrsim \frac{0.03\ M_8^{0.25}}{\epsilon} \;.
\label{eq:lambda}
\end{equation}
This follows from the use of Equations (3), (4), (6) and (7), with
the appropriate definition of $r_{circ}$ in Equation (2).
It is clear that this condition is met for reasonable values of 
$\epsilon$ and $\lambda$ (c.f., Sec.~\ref{sub:AGC}).
Because $\delta_{Fe} \sigma_K \approx \sigma_T$, the flow will then
also be optically thin to Compton scattering.

\subsection{Cloud Formation}
\label{sub:cloudform}

In order for clouds to form, the cooling time, $t_{cool}$, of the shocked gas
must be less than the dynamical time, $t_{dyn}$, for the gas to be transported
through the shock region;  i.e.,  $t_{cool} < t_{dyn}$. The cooling time can be
calculated numerically (c.f. Eq.~\ref{eq:t_{cool}} of the Appendix) if the
initial temperature of the shocked gas is known.  Assuming that the shock converts
the ordered velocity of the flow into random (thermal) motions, the initial
temperature should be $T_s \approx m_H \Delta (v^2) / 3 k$, where $\Delta (v^2)$
is the change in the square of the velocity across the shock.

We assume that the pre-shock conditions are those of the wind; i.e., 
$n_w$ and $v_w$ are used as the pre-shock density and velocity, respectively.
At a strong shock, we have 
\begin{equation}
v_s^{(n)} = v_w^{(n)}/4\;,\qquad v_s^{(t)} = v_w^{(t)}\;,\qquad n_s v_s^{(n)} = n_w v_w^{(n)}\;,
\label{eq:wshock}
\end{equation}
where the superscripts $(n)$ and $(t)$ refer to the normal and tangential velocity
components, respectively, relative to the shock front and $v_s$ is the velocity 
of the shocked gas.  In bow shocks, most of the kinetic energy of the incident 
flow is dissipated, so $\Delta (v^2) \simeq v_w^2$.  For shocks between obliquely 
incident gas flows, it is the component of the wind velocity normal to the shock that
is converted into thermal energy, so $\Delta (v^2) \simeq v_w^{(n)\,2}$.

\subsection{Physical Properties of the Cooled Gas}
\label{sub:PP}

As the shocked gas cools, it clumps to form clouds.  Their physical characteristics,
such as the number density $n_c$, the ionization parameter $U$, and the column depth 
$N_H$ through each clump, all contribute to a determination of the line emissivity 
using photoionization codes such as CLOUDY (\cite{GJF96}).
Assuming isobaric cooling occurs, the density of the cooled gas is
\begin{equation}
n_c = \left( \frac{T_s}{T_{eq}} \right) n_s\;.
\label{eq:n_cloud}
\end{equation}
The ionization parameter, $U$, is then determined directly from the luminous ionizing 
flux, the SED, and $n_c$.  Finally, the column density of a cloud is given by $N_H 
\simeq n_c\; l_c$, where $l_c$ is the cloud size.

In the PD85 model, the maximum cloud size is set by the coherence length,
$l_{coherence} \leq t_{cool}\ c_s$, where $c_s$ is the sound speed in the shocked gas.
If the cooling is isobaric and steady, $l_c = \left(T_{eq} / T_s \right)^{1/3}l_{coherence}$.
It seems that this model is overly optimistic, however, as turbulent mixing is likely 
to be very important in any shock.  Random motions of the turbulent fluid will disrupt 
coherence within the cooling gas; the maximum cloud size is then dictated by the 
smallest scale at which turbulence persists.  Unfortunately, this scale is not 
specified by our simple model, so $N_H$ remains relatively undetermined.

\subsection{Cloud Confinement}

Krolik, McKee, \& Tartar (1981) first proposed the co-existence of cool,
dense clouds (the source of the broad emission lines) confined by a hot,
rarefied medium.  They showed that it was possible, under the right spectral
conditions, to have the two phases in pressure equilibrium (i.e., to have the same
value of $\Xi$) and yet have vastly different temperatures.
Unfortunately, two stable states can only co-exist in very hard AGN spectra,
with $T_C \gtrsim 10^8$~K.  Most AGN spectra are much softer than this, 
effectively ruling out the two-phase pressure equilibrium condition (\cite{ACF86}).  
Dense clouds emerging from the high-pressure shock environment will rapidly 
expand into the ambient flow at their sound speed (e.g., \cite{CSR95}).  
Therefore, unless some other confinement mechanism is
introduced, such as a magnetic field (e.g., \cite{EBS92}), the clouds produced
within a shock are likely to survive only within the shock itself.
As a result, the cloud motions are dictated by the shock motions, which are 
in turn dictated by the wind flow.

\section{RESULTS}

The condition for cooling, i.e., $t_{cool} < t_{dyn}$, sets a minimum length 
scale for the shock region.  Shocked gas in regions smaller than this size will 
simply flow out of the region before it has time to cool.  From the discussion in 
Sec.~\ref{sub:cloudform}, we note that the dynamical time can be expressed as 
$t_{dyn} \approx d_s / v_s^{(n)}$, where $d_s$ is the size of the shock region.
Our requirement for the minimum shock size is therefore
\begin{equation}
d_s > v_s^{(n)}\ t_{cool} \ . 
\label{eq:d_s_min}
\end{equation}

\begin{figure}
\figurenum{1}
\plottwo{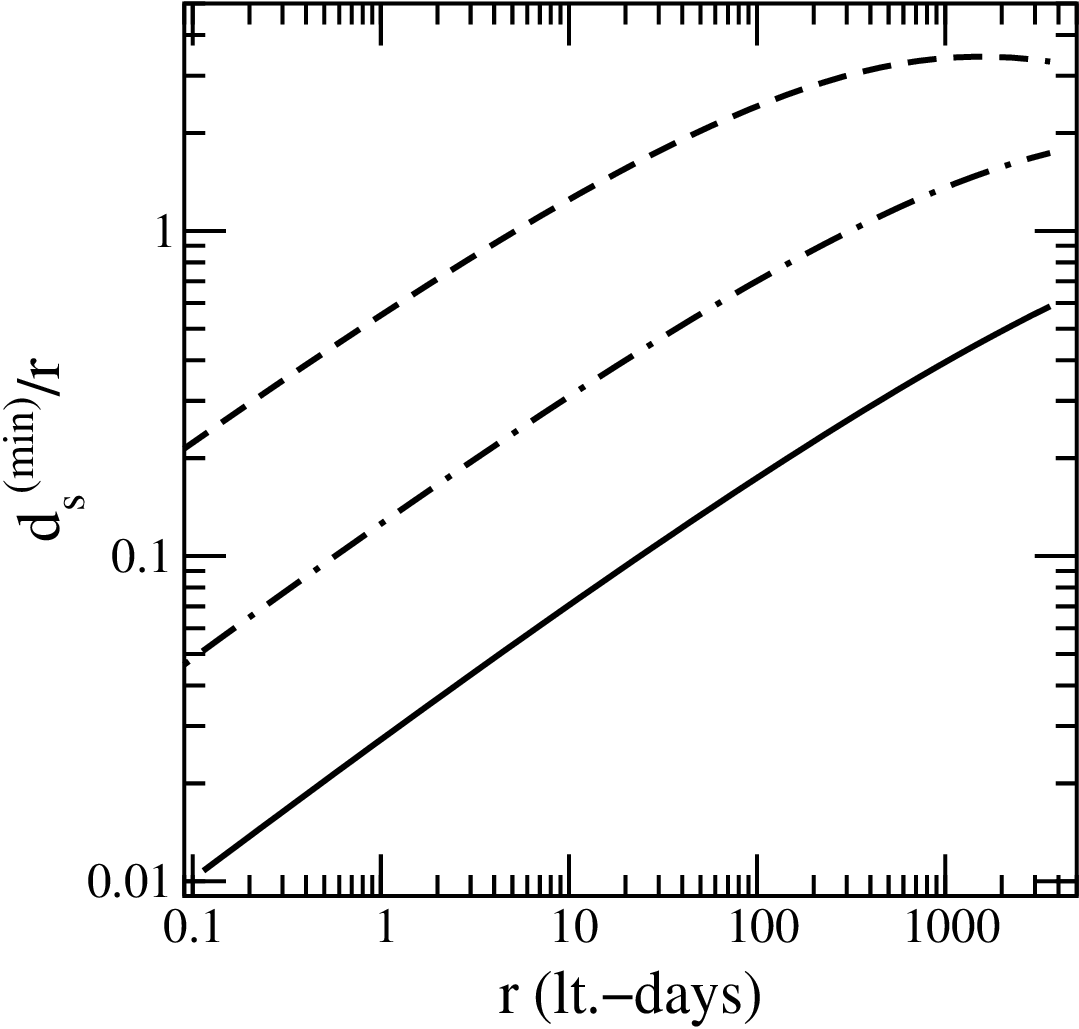}{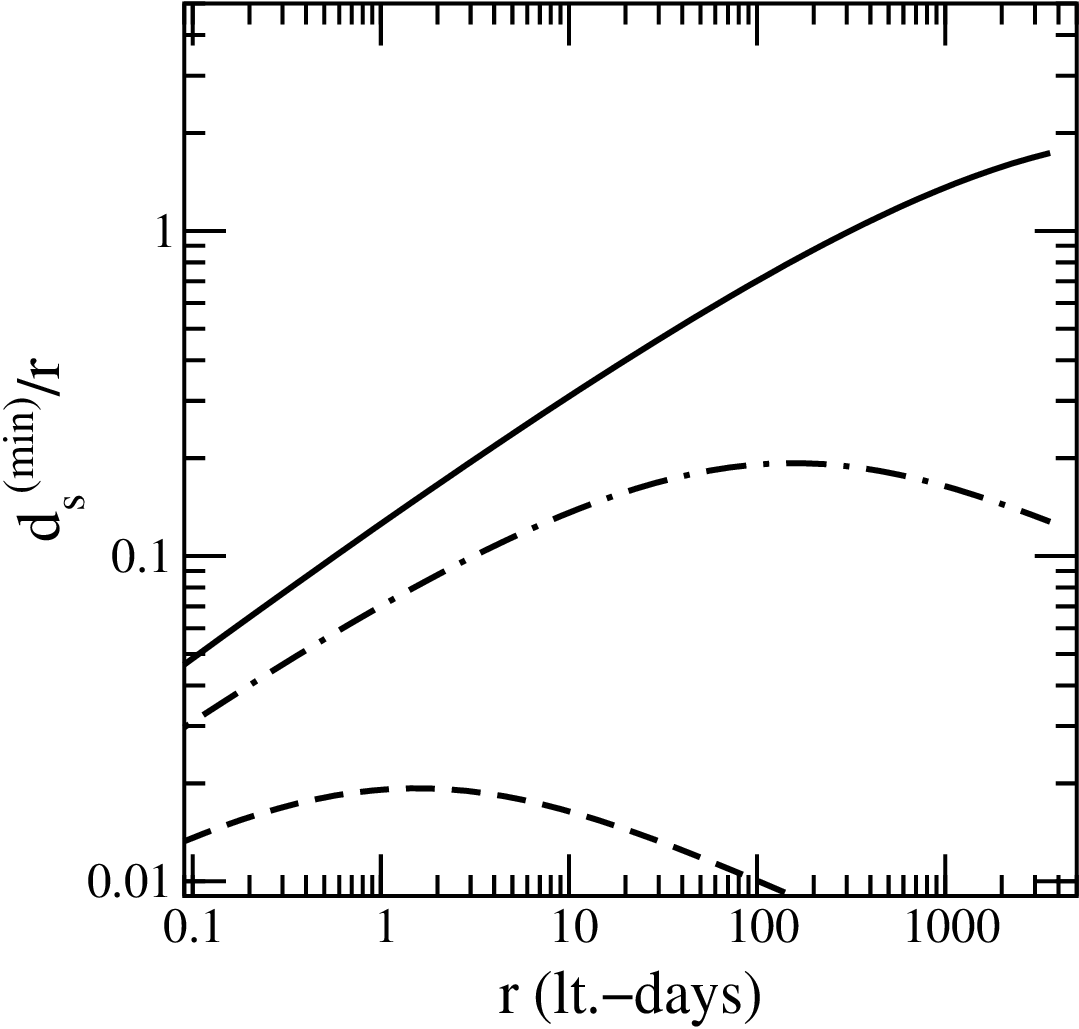}
\caption{The minimum shock length scale required to allow clumping due to radiative cooling.  
Left panel: $\epsilon = 0.1$; $T_7 = 1$; and $M_8 = 0.1$ (dashed), 1.0 (dot-dashed), 10.0 (solid).  
Right panel: $M_8 = 1$; $T_7 = 1$; and $\epsilon = 0.001$ (dashed), 0.01 (dot-dashed), 
0.1 (solid).}
\label{fig:fig1}
\end{figure}

\begin{figure}
\figurenum{2}
\plotone{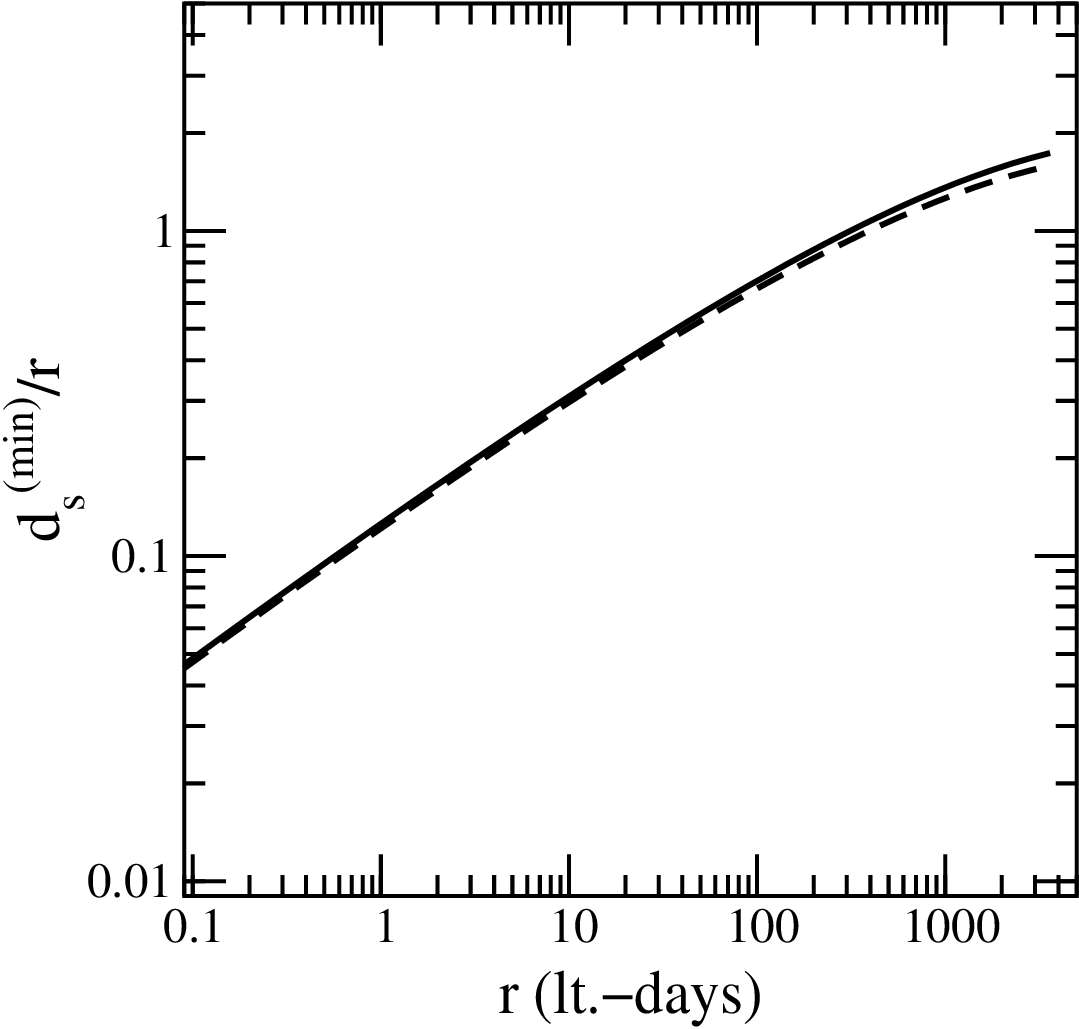}
\caption{The minimum shock length scale required to allow clumping due to radiative cooling,
with a dependence on $T_C$.  $M_8 = 1$; $\epsilon = 0.1$; and $T_7 = 0.1$ (dashed) and 
1 (solid).}
\label{fig:fig2}
\end{figure}

In Figures \ref{fig:fig1} and \ref{fig:fig2}, we plot the minimum value of $d_s$ 
for a range of values in the parameters $M$, $\epsilon$, and $T_C$.
We have here set $v_s^{(n)} = v_w/4$, an upper limit that occurs when the 
colliding winds are incident normally.  Our $d_s^{(min)}$ estimates are therefore 
rather conservative; for obliquely incident winds, smaller shock regions may suffice.
It is reasonable to assume that shocks are possible sites for cloud formation only 
if $d_s^{(min)} \leq r$.

Figure~\ref{fig:fig1} illustrates the effect of varying the central mass $M$ (left) 
and accretion efficiency $\epsilon$ (right).  Increasing $M$ has the effect of 
decreasing $d_s^{(min)}$ at any given radius, thereby extending the plausible 
cloud production region to larger radii.  This is because both the luminosity 
(via Eq.~\ref{eq:L_bol2}) and wind density (via Eq. \ref{eq:L_bol}) increase with $M$.
With Compton cooling being proportional to the luminosity, and bremsstrahlung 
cooling (per particle) being proportional to the density (c.f. Eq.~\ref{eq:F_c}), 
the value of $t_{cool}$ becomes smaller.  Decreasing the value of $\epsilon$ also 
decreases $t_{cool}$; smaller values of $\epsilon$ mean higher values of $n_w$ for 
any given luminosity (Eq. \ref{eq:L_bol}), thereby enhancing the bremsstrahlung cooling rate.

Figure~\ref{fig:fig2} shows the effect of varying the Compton equilibrium temperature $T_C$.
Plotted are the $d_s^{(min)}$ curves for $T_7 = 0.1$ and $1$ with fixed $M$ and $\epsilon$.
There is no great difference between the cooling times, with the cooling occuring 
slightly faster in the case of the softer continuum spectrum.  Note that cooling to 
$T \sim 10^4$~K does \emph{not} occur for $T_7 = 10$.  When the temperature of the 
cooling gas drops below $T_C$, the Compton heating/cooling term in Eq. (\ref{eq:F_c}) 
changes sign; in order for cooling to continue, the bremsstrahlung term must dominate 
at $T \lesssim T_C$.  In the high $T_C$ case, the gas has not cooled sufficiently for 
the (density-dependent) bremsstahlung term to dominate, so the gas simply equilibrates 
at $T_{eq} \simeq T_C$.

\begin{figure}
\figurenum{3}
\plotone{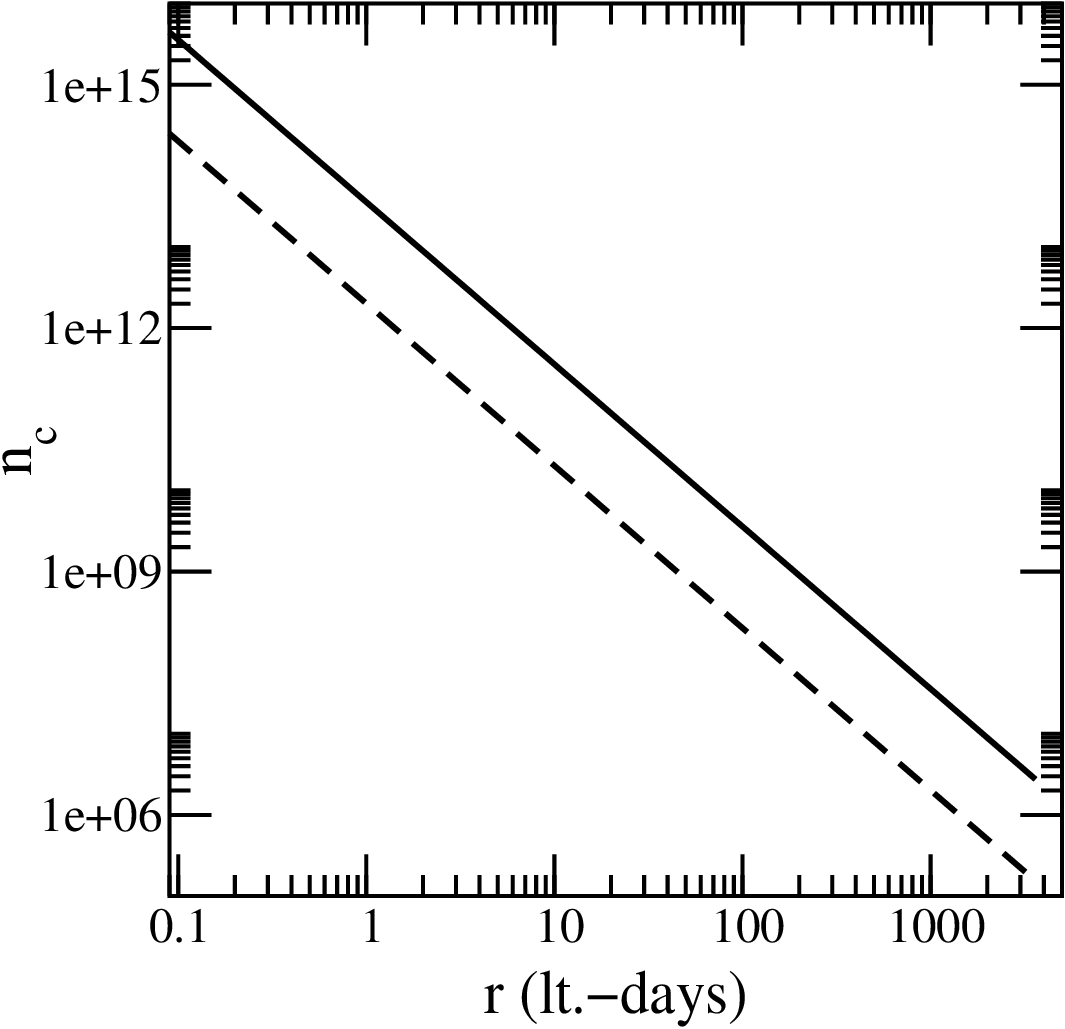}
\caption{Density of shock-produced clouds.  Solid: $M_8 = 1$; $\epsilon = 0.1$ and $0.01$; 
$T_7 = 1$.  Dashed: $M_8 = 0.1$; $\epsilon = 0.1$ and $0.01$; $T_7 = 1$.}
\label{fig:fig3}
\end{figure}

In Figure~\ref{fig:fig3}, we plot the density of the cooled (cloud) gas for a range 
of parameter values.  Note that the gas displays the range in densities over radii 
inferred for the BLR.  We find that the density at any given radius increases with 
$M$, but is insensitive to the values of $\epsilon$ and $T_C$.  Note also that the 
density increases with decreasing $r$; this is consistent with the results of 
modeling the BLR using photoionization codes (e.g., \cite{KN99}).

\section{MODELS OF SHOCK-FORMED CLOUDS}

\subsection{Stellar Wind Bow Shock Model}
\label{sec:SW}

We next study a sample of shock producing mechanisms with the goal of determining 
plausible shock sites for cloud production.  Let us begin by first considering the 
PD85 model, but now with an inflow (due to the accretion of ambient gas onto the
central engine) to act as the agent of interaction with the winds from stars 
embedded within (although not co-moving with) this plasma; this is in contrast
with the outflow assumed by these authors.  In this picture, broad line clouds are 
produced within the bow shocks surrounding the stellar wind sources.

In this case, the size of the shock is determined by the stand-off distance 
(\cite{JJP85}), which gives \begin{equation}
d_s^{(sw)} \simeq 3.1 \times 10^{29} \left( \frac{\dot{E}_{36}}{n_w v_w^2 v_o} 
\right)^{1/2} \ \ {\rm [cm]} \;,
\label{eq:d_s}
\end{equation}
where $\dot{E}_{36}$ is the kinetic energy outflow rate in the stellar wind in units
of $10^{36} \;\mathrm{erg}\;{\mathrm{s}}^{-1}$ and $v_o$ is the outflow velocity.

\begin{figure}
\figurenum{4}
\plotone{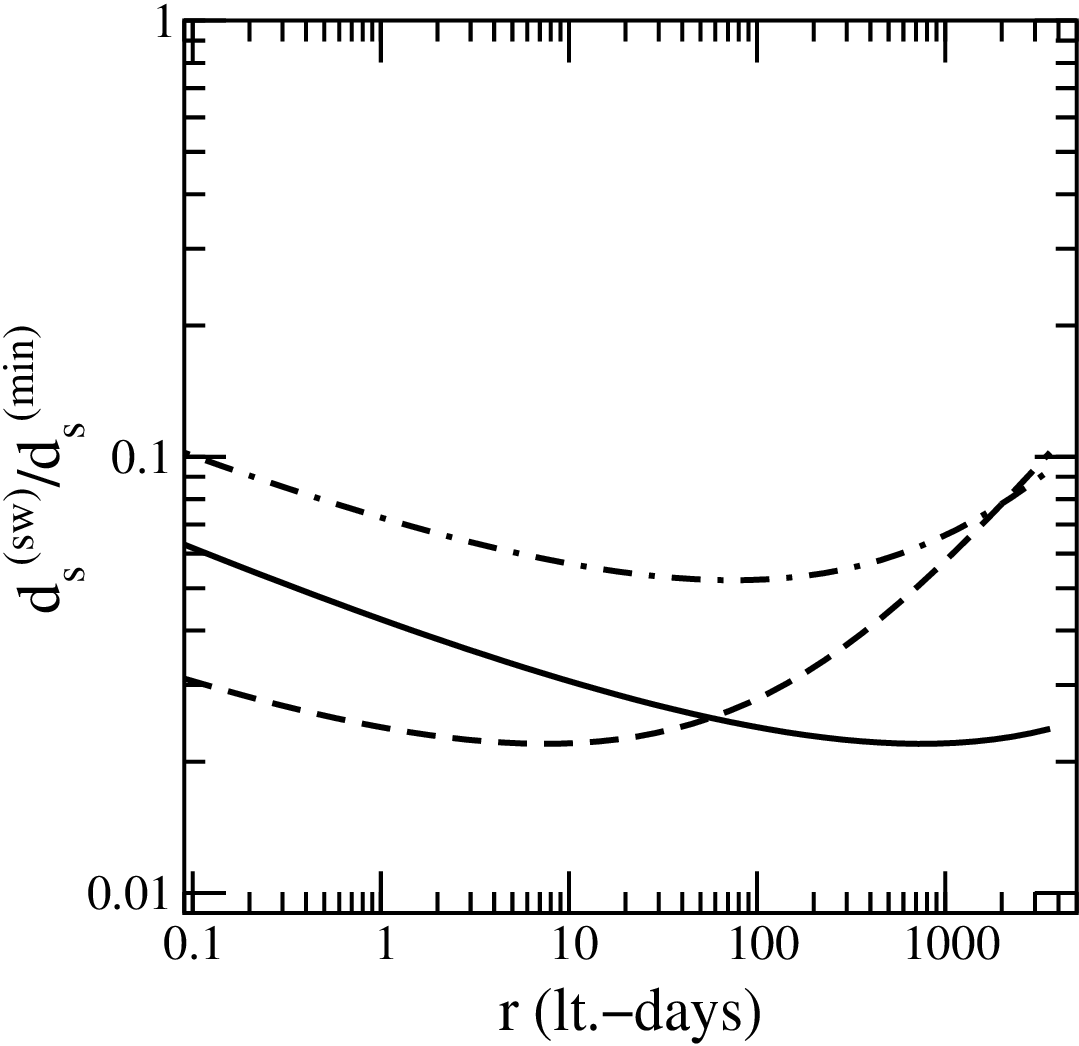}
\caption{Ratio of stellar wind shock size to minimum shock length scale required for
radiative cooling to form clouds.  Solid: $M_8 = 1$, $\epsilon = 0.1$, $T_7 = 1$.  
Dashed: $M_8 = 1$, $\epsilon = 0.01$, $T_7 = 1$.  Dot-dashed: $M_8 = 10$, $\epsilon 
= 0.1$, $T_7 = 1$.}
\label{fig:fig4}
\end{figure}

In Figure~\ref{fig:fig4}, we plot $d_s^{(sw)}/d_s^{(min)}$ using the fiducial 
values $v_o \approx 2,000 \;\mathrm{km}\;{\mathrm{s}}^{-1}$ and $\dot{E}_{36} 
\approx 100$, typical for W-R stars (although these are probably upper limits 
for a typical stellar population).  It can be seen that the size of these shocks 
is probably too small for these to be viable sites for cloud production via radiative 
cooling, thus confirming the PD85 result.

\subsection{Bondi-Hoyle Accretion Shock Model}
\label{sec:BH}

In the Bondi-Hoyle accretion process, a bow shock forms around the black hole 
when it accretes from a rather uniform, laminar flow.  The length scale of the shock 
is roughly the accretion radius itself, i.e., $d_s^{(BH)} \approx (0.1-1)\,R_A 
\simeq (0.1-1)\,r$ (see Sec.~\ref{sub:AGC}).  We have already seen in Figs. 
(\ref{fig:fig1}) and (\ref{fig:fig2}) that the requirement $d_s^{(min)} < r$ 
can be met for a wide variety of parameters over the range of relevant radii.
Therefore, Bondi-Hoyle shocks around the central mass concentration
are plausible sites for cloud production.

However, it should be noted that $R_A$ is unlikely to remain the only relevant 
scale as a Bondi-Hoyle shock is likely to break up into smaller scale shocks in a 
realistic (unsteady) flow.  This would have the effect of reducing $t_{dyn}$.
Production of BLR clouds by this mechanism is therefore dependent on the stability 
of the large scale shock structure.

An important signature of the highly ordered flow around a Bondi-Hoyle shock 
would be a rather narrow line emission profile whose overall redshift is 
dependent on viewing angle.  This is because the clouds flowing along a shock have 
roughly parallel velocities.  In addition, the Bondi-Hoyle shock does not provide 
a sufficiently broad distribution of cloud properties inferred for an extended BLR.  
It is therefore unlikely that a single Bondi-Hoyle accretion shock could produce 
the broad line profiles seen in AGN spectra.

\subsection{Wind Collision and Turbulent Accretion Shock Model}
\label{sec:T}

The final source of BLR clouds we consider here is shocks produced by large-scale 
wind collisions and turbulence within the overall flow.
The motivation for this is that realistic 3D simulations of the accretion
onto a massive nucleus from a distribution of wind sources (Coker
\& Melia 1997) indicate that a single Bondi-Hoyle bow shock is difficult to form
or maitain.  Instead, the stellar wind-wind collisions produce an array of 
shock segments and a consequent turbulent inflow towards the black hole.
In this picture, clouds are produced continually throughout the extended BLR,
so we avoid the problem of having to confine long-lived clouds; instead clouds 
that evaporate upon leaving the shock region are continually replaced by newly 
formed clouds at other locations within the inflow.

The shock regions must be large to allow cooling to occur (c.f., Figs.~\ref{fig:fig1} 
and \ref{fig:fig2}), but these are readily obtainable for a realistic flow.
Because the shocks, and therefore the clouds themselves, are embedded within the
overall accretion pattern, the velocity of the clouds is roughly equal to that of 
the captured wind; i.e., $v_c \simeq v_w$.  Clouds that move at nontrivial velocities
relative to the surrounding medium are subjected to disruption via Rayleigh-Taylor
instabilities (\cite{MF87}).  The winds, and therefore the clouds, display a 
$v(r) \propto r^{-1/2}$ velocity field fully consistent with, e.g., the 
Peterson \& Wandel (1999) conclusion that the BLR velocity fields in NGC~5548 mimic 
Keplerian motions about a single central mass.  Assuming that a large number of shocks 
exist at different locations within the flow, it should be possible to reproduce 
the observed line profiles.  We have found it quite straightforward to do this
within the context of this model. However, since the actual profile depends on the number 
and location of the shocks, which are not known \emph{a priori}, an actual fitting
such as this is not yet warranted.

\section{CONCLUSIONS}

In this paper, we have considered a broad range of possible gas configurations 
in a wind accreting onto the central black hole, with physical conditions that 
may produce BLR clouds via cooling instabilities within shocks.  We note that 
in order to reproduce the observed line shape in actual sources, the BLR clouds 
cannot all be produced within a single outer region such as a Bondi-Hoyle shock, 
since this does not account for the required range in cloud properties at 
smaller radii.  Instead, we have found that the best scenario involves local 
cloud production throughout the overall accretion flow.
We conclude that a viable model for the formation of the BLR is one in which 
ambient gas surrounding the black hole (e.g., from stellar winds) is captured 
gravitationally and begins its infall with a (specific angular momentum) 
$\lambda$ representative of a flow produced by many wind-wind collisions 
and turbulence rather than a smooth Bondi-Hoyle bow shock.
In this process, the gas eventually circularizes at $r_{circ} \approx 2 \lambda^2 R_S$, 
but by that time all of the BLR clouds have been produced, since at that radius the gas 
presumably settles onto a planar disk.  As such, this picture is distinctly different 
from models in which the clouds are produced within a disk and are then accelerated 
outwards by such means as radiation pressure or magnetic stresses.

We note that these results are consistent with current reverberation studies.
In our model, the inner radius of the BLR is determined by the circularization radius.
Our model predicts that $r_{circ} \approx 2\,\lambda_{10}^2\,M_8$~lt.-days, where 
$\lambda_{10} \equiv \lambda/10$, which is consistent with the inferred inner BLR radius 
of a few light days.  This scenario is also consistent with the differential line 
response seen in most sources (e.g., \cite{KTK95}), with the blue and red wings 
responding fastest to continuum changes before the central peak.  Our picture of the 
BLR has the broad wings formed by rapidly moving gas at small radii and the 
central peak formed by slower moving gas at larger radii.

Finally, we consider the argument that broad line emission cannot be produced by 
discrete clouds (\cite{NA98}).  This reasoning is based on the assumption that 
each cloud has a fixed set of parameters, such as density, thickness, velocity, etc.
In our model, with the clouds continually forming in regions of high turbulence, 
\emph{each} cloud region can display a wide range of properties.  Therefore, we 
suggest that the cross-correlations that appear with fixed cloud properties 
would vanish.  Given the viability of this picture, it now remains to be seen 
whether the vast array of BLR phenomena observed in sources ranging from Seyferts 
to high redshift quasars can be self-consistently accounted for with this single
description.  This is work in progress and the results will be
reported elsewhere.

\bigskip

\section{ACKNOWLEDGEMENTS}
The authors would like to thank Amri Wandel for his helpful review.
This work was supported by an NSF Graduate Fellowship at the
University of Arizona, by a Sir Thomas Lyle Fellowship and a
Miegunyah Fellowship for distinguished overseas visitors at 
the University of Melbourne, and by NASA grant NAG58239.

\appendix

\section{Calculation of the Cooling Time Scale}

Following Krolik, McKee, \& Tartar (1981) and Perry \& Dyson (1985), we calculate the cooling time scale for a shocked gas to cool from an initial temperature, $T_i$, to the equilibrium temperature, $T_f$, using the expression
\begin{equation}
        t_{cool} = \int_{T_f}^{T_i} \frac{d T}{F_c (T)}\;,
        \label{eq:t_{cool}}
\end{equation}
where $F_c (T)$ is the net cooling rate.
From Matthews \& Ferland (1987), we have that
\begin{equation}
        F_c (T) = \frac{4 \sigma_T}{3 m_e c^2} \frac{L_{bol}}{4 \pi r^2} (T - T_C) + \frac{n \lambda_B T^{1/2}}{3 k} \;.
        \label{eq:F_c}
\end{equation}
The first term is due to Compton heating/cooling processes, and the second term is due to bremsstrahlung (free-free) emission.

Equation~(\ref{eq:F_c}) is valid for $T \gtrsim 10^5\ \rm K$.
Below this temperature, collisional and radiative transitions in the plasma cause very rapid cooling. Therefore, $T_f = 10^5\ \rm K$ is taken as the lower limit of integration for the cooling processes described in this paper.

\clearpage

\clearpage

\clearpage

\clearpage

\end{document}